
\input harvmac

\def  \TF {\tilde F}
\def \tp {{\tilde \p}}

\def\bd {{\bar \del}}\def \ra {\rightarrow}

\def \EE {{\bar {\cal  E}}}
\def \ll {{\bar \l}}
\def \J {{\bar J}}
\def \M {{\cal M}}
\def \ao {{\a_1}}
\def \am  {{-\a_1}}

\def \tu {\tilde u }
\def \tv {\tilde v}
\def \sms {sigma models \ }
\def \ca
\def \P {\Phi}

 \def \k1 {{1\over
k}} \def \bh { {\bar h} } \def \ov { \over }
  \def \B { { \bar B }}

\def \ra {\rightarrow}

\def \a {\alpha}
\def \b {\beta}

\def \Tr {{\ \rm Tr \ }}

\def \ln {{\rm \ ln \  }}
\def \det {{ \rm det \ }}

\def \l {\lambda}
\def \1p {{1\over  \pi }}
\def \2p {{{1\over  2\pi }}}
\def \4p {{ {1\over 4 \pi }}}
\def \8p {{{1\over 8 \pi }}}
\def \P^* { P^{\dag } }
\def \p {\phi}
\def \M {{\cal M}}
\def \m {\mu }
\def  \n {\nu}
\def \ep {\epsilon}
\def\g {\gamma}
\def \r {\rho}
\def \k {\kappa }
\def \d {\delta}

\def \s {\sigma}

\def \fourth {{\textstyle{1\over 4}}}

\def \eq#1 {\eqno {(#1)}}
\def \sm {sigma model\ }\def \B  {{ \tilde B }}

\def \bd  {{ \bar \del }}

\def \M {{\cal M}}

\def \B  {{ \tilde B }}

\def \bd  { \bar \del }

\def \ov {\over }

\def \A  { {\bar A} }

\def \bu {{\bar w}}


\def \p {\phi}
\def \ep {\epsilon}
\def \s {\sigma}

\def \r {\rho}
\def \d {\delta}
\def \l {\lambda}
\def \m {\mu}
\def \g {\gamma}
\def  \n {\nu}

\def \fourth {{1\over 4}}

\def \B {{\bar B}}

\def \rank {{\rm rank\ }}
\def \J {\bar J }
\def \P {\Phi }
\baselineskip8pt
\Title{\vbox
{\baselineskip8pt{\hbox{Imperial/TP/93-94/17}}{\hbox{PRA-HEP 94/1}  }
{\hbox{hep-th/9402120}} }}  {\vbox{\centerline
{ Exact  four dimensional string solutions  }
\centerline
{and Toda-like  sigma models}\centerline{  from  `null-gauged' WZNW theories }
}}
\vskip -30 true pt
\centerline  {   C. Klim\v c\'\i k\footnote{$^{*}$}{\baselineskip5pt
e-mail:  presov@cspuni12.bitnet} }
\vskip 2pt
\centerline {\it  Theory Division, Nuclear Centre, Charles University,  }
\centerline {\it   180 00 Prague 8, Czech Republic}
\vskip 2pt
\centerline {and}
\vskip 3pt
\centerline{   A.A. Tseytlin\footnote{$^{**}$}{\baselineskip5pt
On leave  from Lebedev  Physics
Institute, Moscow, Russia.
e-mail:  tseytlin@ic.ac.uk} }
\vskip 2pt
\centerline {\it   Theoretical Physics  Group, Blackett Laboratory, Imperial
College}
\centerline {\it London SW7 2BZ, U.K.}
\vskip 4pt

\baselineskip5pt
\noindent
We  construct  a new class of exact string solutions with a four dimensional
target space metric  of  signature ($-,+,+,+$)  by gauging the independent left
and right  nilpotent subgroups with `null' generators of WZNW models for rank 2
non-compact groups $G$.
The `null' property of the generators (${\rm Tr }(N_n N_m)=0$) implies the
consistency of the gauging and the absence of $\a'$-corrections to the
semiclassical
backgrounds obtained from the  gauged WZNW models.
In the case of the maximally non-compact groups ($G= SL(3), SO(2,2), SO(2,3),
G_2$)
the  construction   corresponds to   gauging  some of
the subgroups generated by
 the nilpotent `step' operators in
the Gauss decomposition.
The rank 2 case is a particular example of a general construction  leading to
conformal backgrounds
with one time-like direction.
The conformal theories  obtained by integrating out the gauge field can be
considered as sigma model analogs of Toda models (their classical
equations of motion are equivalent to Toda model equations).
The procedure of `null gauging'  applies also to other  non-compact  groups.
As an example,  we consider the  gauging
of $SO(1,3)$  where the   resulting metric  has the signature
 ($-,-,+,+$) but admits  two   analytic continuations with Minkowski signature.
The   backgrounds  we find  have `2+2' structure with two null Killing vectors.
Their  dual counterparts have one covariantly constant  null  Killing vector,
i.e.
 are of  `plane-wave' type  (with    metric and dilaton depending only on
transverse spatial coordinates) and  also represent exact string solutions.

\Date {{February 1994}
 }


\noblackbox
\overfullrule=0pt
\baselineskip 20pt plus 2pt minus 2pt

\def\np {  Nucl. Phys. }
\def \pl { Phys. Lett. }
\def \mpl { Mod. Phys. Lett. }
\def \prl { Phys. Rev. Lett. }
\def \pr  { Phys. Rev. }

\def \cmp { Commun. Math. Phys. }
\def \ijmp { Int. J. Mod. Phys. }

\lref \anton { I. Antoniadis, C. Bachas, J. Ellis and D.V. Nanopoulos,
\pl B211(1988)393. }
\lref \horav{P. Ho\v rava, \pl B278(1992)101.   }
\lref\gersh {  D. Gershon,  \pr D49(1194)999;  hep-th/9210160,
9311122. }

\lref \bcr {K. Bardakci, M. Crescimanno and E. Rabinovici, \np
B344(1990)344. }
\lref \wiit { E. Witten, \pr D44(1991)314.}
\lref \gwz {      K. Bardakci, E. Rabinovici and
B. S\"aring, \np B299(1988)157;
 K. Gawedzki and A. Kupiainen, \pl B215(1988)119;
\np B320(1989)625. }

\lref \karabali { D. Karabali, Q-Han Park, H.J.
Schnitzer and
Z. Yang, \pl B216(1989)307;  D. Karabali and H.J. Schnitzer, \np B329(1990)649.
}
\lref \polwig { A.M. Polyakov and P.B. Wiegman,  \pl
B141(1984)223.  }

 \lref \kumar {A. Kumar,
\pl B293(1992)49; D. Gershon, preprint TAUP-2005-92.}
\lref  \hussen {  S. Hussan and A. Sen,  \np B405(1993)143. }
\lref \kiri {E. Kiritsis, \np B405(1993)109. }
\lref \alv { E. Alvarez, L. Alvarez-Gaum\'e, J. Barb\'on and Y. Lozano,
preprint CERN-TH.6991/93.}

\lref  \rocver { M. Ro\v cek and E. Verlinde, \np B373(1992)630.}

\lref \brink{ H.W. Brinkmann, Math. Ann. 94(1925)119.}
\lref \guv {R. G\"uven, Phys. Lett. B191(1987)275.}

\lref \amkl { D. Amati and C. Klim\v c\'\i k, \pl B219(1989)443.}

\lref \hor { G. Horowitz and A.R. Steif, Phys.Rev.Lett. 64(1990)260 ;
Phys.Rev. D42(1990)1950.}

\lref \horr {G. Horowitz, in: {\it Proceedings
of  Strings '90},
College Station, Texas, March 1990 (World Scientific,1991).}

\lref \rudd { R.E. Rudd, \np B352(1991)489 .}

\lref \tsnul { A.A. Tseytlin, \np B390(1993)153.}
\lref \tsnull { A.A. Tseytlin, \pl B288(1992)279; \pr D47(1993)3421.}

\lref \dunu { G. Horowitz and A.A. Steif, \pl B250(1990)49;
 E. Smith and J. Polchinski, \pl B263(1991)59;}

\lref\horhorst  {J. Horne, G. Horowitz and A. Steif, \prl 68(1992)568;
G. Horowitz, in:  {\it  Proc. of the 1992 Trieste Spring School on String
theory and Quantum Gravity},
preprint UCSBTH-92-32, hep-th/9210119.}
\lref \horwel {
 G. Horowitz and D.L. Welch, \prl 71(1993)328.
}

\lref \kala { N. Kaloper, \pr D48(1993)2598.}

\lref \sfetsos { K. Sfetsos, preprint USC-93/HEP-S1, hep-th/9305074.}
 \lref \busc { T.H. Buscher, \pl B194(1987)59 ; \pl B201(1988)466.}
\lref \pan  { J. Panvel, \pl B284(1992)50. }

\lref \mye {  R. Myers, \pl B199(1987)371;
    I. Antoniadis, C. Bachas, J. Ellis, D. Nanopoulos,
\pl B211(1988)393;
 \np B328(1989)115. }
\lref \givpassq { A. Giveon and A. Pasquinucci, \pl B294(1992)162. }
\lref \GK {A. Giveon and E. Kiritsis,  preprint CERN-TH.6816/93,
RIP-149-93.}
\lref \givroc {A. Giveon and M. Ro\v{c}ek, Nucl. Phys. B380(1992)128.}
\lref \giv {  A. Giveon, Mod. Phys. Lett. A6(1991)2843;  R. Dijkgraaf, H.
Verlinde and
E. Verlinde, \np B371(1992)269; I. Bars, preprint USC-91-HEP-B3;
 E. Kiritsis, \mpl A6(1991)2871. }
\lref \dvv {  R. Dijkgraaf, H.
Verlinde and
E. Verlinde, \np B371(1992)269. }

\lref \GRV {A. Giveon, E. Rabinovici and G. Veneziano, Nucl. Phys.
B322(1989)167;
A. Shapere and F. Wilczek, \np B320(1989)669.}
\lref \GMR {A. Giveon, N. Malkin and E. Rabinovici, Phys. Lett. B238(1990)57.}
\lref \Ve    { K.M. Meissner and G. Veneziano, \pl B267(1991)33;
M. Gasperini, J. Maharana and G. Veneziano, \pl B272(1991)277; \pl
B296(1992)51.}
\lref \sen {A. Sen,  \pl  B271(1991)295; \pl B274(1991)34.}

\lref \VV    {  G. Veneziano, \pl B265(1991)287.}

\lref  \horne { J.H. Horne and G.T. Horowitz, \np B368(1992)444.}

\lref \koki { K. Kounnas and  E. Kiritsis, preprint CERN-TH.7059/93;
hep-th/9310202. }
\lref \napwit { C. Nappi and E. Witten, \prl 71(1993)3751.}

\lref \napwi { C. Nappi and E. Witten, \pl B293(1992)309.}

\lref \tsmpl {A.A. Tseytlin, \mpl A6(1991)1721.}

\lref \tsbh {A.A. Tseytlin,  preprint CERN-TH.6970/93; hep-th/9308042.}

\lref \nsw {  K.S. Narain, M.H. Sarmadi and E. Witten, \np B279(1987)369. }
\lref \call {C.G. Callan, D. Friedan, E. Martinec and M.J. Perry,
Nucl. Phys. B262 (1985)593 ;
E. S. Fradkin and A. A. Tseytlin, \np B261(1985)1;
  A.A. Tseytlin, \pl B178(1986)349.}

\lref \ban {M. Banados, C. Teitelboim and J. Zanelli, \prl 69(1992)1849.}

\lref \STT {K. Sfetsos and A.A. Tseytlin, preprint CERN-TH.6969/93,
hep-th/9310159.}
\lref \givkir {A. Giveon and E. Kiritsis,  preprint CERN-TH.6816/93;
hep-th/9303016. }

\lref \cec { S. Cecotti, S. Ferrara and L. Girardello, \np B308(1988)436. }

\lref \tsdu { A.A. Tseytlin, \np B350(1991)395.  }

\lref \wit {E. Witten, Commun. Math. Phys. 92(1984)455 ;
E. Braaten, T.L. Curtright and C.K. Zachos, \np B260(1985)630.}

\lref \sftss{ K. Sfetsos and A.A. Tseytlin, unpublished (1994). }
\lref \witte { E. Witten, \cmp 144(1992)189.}
\lref \koulust { E. Kiritsis, C. Kounnas and D. L\"ust, preprint
CERN-TH.6975/93,
hep-th/9308124.}
\lref \kou { C. Kounnas, preprint CERN-TH.6799/93,
hep-th/9304102;
I. Antoniadis, S. Ferrara and C. Kounnas, preprint CERN-TH.7148/94,
hep-th/9402073.}

\lref \oli { D. I. Olive, E. Rabinovici and A. Schwimmer, \pl B321(1994)361.}
\lref \sfet {K. Sfetsos, \pl B324(1994)335; preprints  THU- 93/31, 94/01;
hep-th/9311093,
 9402031.}
\lref \moham { N. Mohammedi, \pl B325(1994)371.}

\lref\ginspqu   { P. Ginsparg and F. Quevedo, \np B385(1992)527.}
\lref\kounn { C. Kounnas and D. L\" ust, \pl B289(1992)56.}
\lref \gidd { S.B. Giddings, J. Polchinski
and A. Strominger, \pr D48(1993)5784.}

\lref  \guv {
R. G\"uven, Phys. Lett. B191(1987)275.}

\lref \amkl { D. Amati and C. Klim\v c\'\i k, \pl B219(1989)443.}
\lref  \hst { G. Horowitz and A.R. Steif, Phys.Rev.Lett. 64(1990)260;
Phys.Rev. D42(1990)1950.}
\lref \petrov { A.Z. Petrov, Einstein Spaces (Pergamon, N.Y. , 1969).}
\lref \figue {J.M. Figueroa-O'Farrill and S. Stanciu, preprint QMW-PH-94-2,
 hep-th/9402035.}
\lref \ira { M. Alimohammadi, F. Ardalan and H. Arfaei, preprint BONN-HE-93-12,
 hep-th/9304024.}
\lref \ind { A. Kumar and S. Mahapatra, preprint IP/BBSR/94-02,
hep-th/9401098.}
\lref \oraif {P. Forg\'acs, A. Wipf, J. Balog, L. Feh\'er and L.
O'Raifeartaigh,
\pl B227(1989)214;  J. Balog, L. Feh\'er,  L. O'Raifeartaigh,
 P. Forg\'acs
and A. Wipf,  Ann. Phys. 203(1990)76;
L. O'Raifeartaigh,
 P. Ruelle and I. Tsutsui, \pl B258(1991)359.}
\lref \sen {A. Sen, \pl B274(1992)34.}

\lref \witt { E. Witten, \cmp 92(1984)455.}
\lref \gwzw  { P. Di Vecchia and P. Rossi, \pl  B140(1984)344;
 P. Di Vecchia, B. Durhuus  and J. Petersen, \pl  B144(1984)245.}

\lref\barsf { I. Bars and K. Sfetsos, \pl B277(1992)269;
 \pr D46(1992)4495, 4510;
I. Bars, preprint USC-93/HEP-B3,  hep-th/9309042. }
\lref\bs { I. Bars and K. Sfetsos,  \pr D48(1993)844. }
\lref \ts {A.A.Tseytlin, \np B399(1993)601; \np B411(1994)509.}
\lref \arfa {H. Arfaei and N. Mohammedi, preprint BONN-HE-93-42,
hep-th/9310169.}
\lref \feher {L. Feh\'er,  L. O'Raifeartaigh,  P. Ruelle, I. Tsutsui
and A. Wipf,  Phys. Rep. 222(1992)1. }

\lref \helgas {S. Helgason, Differential Geometry, Lie Groups and Symmetric
Spaces (Academic Press, N.Y., 1978).}
\lref \jack { I. Jack and J. Panvel, preprint LTH-304,  hep-th/9302077.}
\lref \sfts {K. Sfetsos and A.A. Tseytlin, \np B415(1994)116,
hep-th/9308018.}
\lref \jackk { I. Jack, D.R.T. Jones and
 J. Panvel, preprint  LTH-315, hep-th/9308080.}
\lref \tye { S.-W. Chung and S.-H. Tye, \pr D47(1993)4546.}
\lref \giveon { M. Ro\v cek and E. Verlinde, \np B373(1992)630;
 A. Giveon, M. Porrati and E. Rabinovici, preprint RI-1-94,  hep-th/9401139. }

\lref\bakas  {I. Bakas, preprint CERN-TH.7144/94,   hep-th/9402016. }

\lref \hum {J.E. Humphreys, Introduction to Lie Algebras and Representation
Theory
 (Springer, New York 1972) }

\lref \baru { A. Barut and R. Raczka, ``Theory of Group Representations and
Applications" (PWN, Warszawa 1980). }
\lref \zhel {D. Zhelobenko and A. Shtern,  Representations of Lie Groups
(Nauka, Moscow 1983). }

\lref \bal { J. Balog, L. O'Raifeartaigh,  P. For\' gacs
and A. Wipf,  \pl B325(1989)225.
 }
\lref  \barn { I. Bars and D. Nemeschansky, \np B348(1991)89.}
\lref \olive{D.I. Olive, ``Lectures on gauge theories and Lie algebras",
Imperial College preprint (1982).}
\lref \schts{A.S. Schwarz and A.A. Tseytlin, \np B399(1993)691.    }
\lref  \mans { P. Mansfield, \np B222(1983)419.}
\lref \tsey { A.A. Tseytlin, \pl B241(1990)233.}
\lref \ger {A. Gerasimov, A. Morozov, M. Olshanetsky, A. Marshakov and S.
Shatashvili, \ijmp
A5(1990)2495. }
\lref \gervais {J.-L. Gervais and M.V. Saveliev, \pl B286(1992)271.   }
\lref \bil {A. Bilal and J.-L. Gervais, \pl B206(1988)412; \np B318(1989)579. }
\lref \lez {A.N. Leznov and M.V. Saveliev, \cmp 74(1980)111;  89(1983)59.}
\lref \wip {L. O'Raifeartaigh
and A. Wipf,  \pl B251(1989)361. }
\lref \shat {A. Alekseev and S. Shatashvili, \np B323(1989)719.}
\lref \ber  {M. Bershadsky and H. Ooguri, \cmp 126(1989)49.}
\lref \oltu {D. Olive and N. Turok, \np B220(1983)491.}
\lref \klts { C. Klim\v c\'\i k  and A.A. Tseytlin, \pl B323(1994)305;
hep-th/9311012.}
\lref \natd {J.-L.  Gervais, L. O'Raifeartaigh, A.V.  Razumov and M.V.
Saveliev, preprint DIAS-92/27, hep-th/9211088.}
\lref \hots{G. Horowitz and A.A. Tseytlin, to appear}
\lref \berg {E. Bergshoeff, I. Entrop and R. Kallosh, preprint SU-ITP-93-37;
hep-th/9401025.}


\newsec{Introduction}

 In trying to understand   possible implications of string theory for
gravitational physics it is important  to   study
string solutions  with  physical dimension  ($D=4$) and signature
($-,+,+,+$).   With a hope to be able to discuss  issues of singularities and
short distance structure
one is  mostly interested not just  in  solutions of the leading-order
low-energy  string effective equations but in the ones  that  are exact
in $\a'$ and/or which have  an  explicit  conformal field theory
interpretation.
While the leading-order solutions are  numerous, very few  solutions of the
second type  are known.   In addition to the obvious `direct product'
combinations
of low-dimensional solutions (e.g. $R\times SU(2)$ WZNW  \anton) the known
exact  $D=4$ ones include, in particular,
spaces with a covariantly constant null Killing vector
\guv\amkl\hst\rudd\tsnull, `black hole' - type  \horav\gersh\sen\ and
cosmological \ginspqu\kounn\napwi\givpassq\ solutions based on $(G\times
G')/(H\times H')$
gauged WZNW models  (with $G=SL(2,R) ,\  G'= SL(2,R)$ or $SU(2)$
and $H, H' = U(1) $ or $SO(1,1)$),\foot{For other superstring solutions see
also \kou\koulust\ and references there. }
 the $SO(2,3)/SO(1,3)$  model  of \barsf,
a black hole solution of  \gidd\ and
the solution \napwit\  corresponding to the   WZNW model for a non-semisimple
$D=4$ group
 which has a non-degenerate
invariant bilinear form.\foot{Since there is  no dimension 4 simple Lie group
one may try to obtain a $D=4$ solution from a WZNW-type theory either by
considering  $G/H$ gauged models or by using the construction \napwit\oli\sfet\
based on non-semisimple groups.
  It is possible to  check   explicitly \sftss\  that there are no other (in
addition to the algebra in \napwit)
non-trivial $D=4$ solvable Lie algebras (for a classification see \petrov)
which  have a non-degenerate invariant bilinear
form  and thus  could lead to new WZNW-type models  according to  \moham.
The existence of only one $D=4$ non-abelian algebra  with an invariant form  is
explained by a theorem in a recent paper \figue :  since non-semisimple
algebras with invariant forms are obtained by the procedure of ``double
extension" ($g\ra g\oplus h\oplus h^*$) from an algebra  $g$ with an invariant
form, to get a $D=4$ algebra one needs to  start with a $D=2$ algebra as $g$
($h$ must be one-dimensional to get $D=4$). Among the two
 $D=2$  Lie algebras only the abelian one can have a non-degenerate  invariant
form; the corresponding $D=4$ algebra  is the central extension of the
euclidean group in two dimensions, the algebra used in \napwit.}

The aim of this paper is to present some new    exact $D=4$  solutions
which correspond to gauged WZNW models and thus  should have a direct conformal
field theory interpretation.  The starting point will be   what we shall call
`null gauged' WZNW models,
i.e. gauged  WZNW models  \witt\gwzw\gwz\karabali\ based on non-compact
\bal\barn\   groups
 with the generators of the  gauged subgroup being `null' (having zero Killing
scalar products). The  gauged subgroup  will be  thus chosen  to be solvable
(but  need not be nilpotent in general). We shall  generalise the  procedure of
 gauging  WZNW models with nilpotent  subgroups  \oraif\
to obtain  conformal \sms with Minkowski signature.
The resulting  \sms will belong to the following class
$$  S = {1\ov \pi \a'} \int d^2 z
 \big[   \del x^i \bd x_i
   +   F(x)   \del u \bd v  \big]   +  {1\ov 4 \pi  }  \int d^2 z \sqrt
{g^{(2)}}  R^{(2)} \p (x)   \  ,       $$
where the  two functions $F$ and $\p$  will  be explicitly determined (in
Section 2)  by gauging of  the   nilpotent   subgroups   in WZNW models for
rank
$n$ maximally non-compact groups.  $x^i$ ($i=1,...,n$)  will be the linear
combinations of  the coordinates  $r^i\equiv r^{\a_i}$  corresponding to the
simple roots  $\a_i$,
$$\del x^i \bd x_i=  C_{ij}  \del r^i\bd r^j \ , \ \ \a_i\cdot x = K_{ij} r^j \
, \ \
 \  \ K_{ij}\equiv K_{\a_i\a_j}= {2 \a_i \cdot \a_j \ov  |\a_j|^2}  = \ha
|\a_i|^2  C_{ij} \ , $$
where $ K_{ij}$ is the  $n\times n$ Cartan matrix. We shall  find that
$$ F=  {1\ov   \sum_{i=1}^n \ep_i {\rm e}^{  \a_i\cdot x  } }\ , \  \ \ \ \
\p =  \ha \sum_{s=1}^m  \a_s\cdot x  -  \ha
 \ln   \big(\sum_{i=1}^n  \ep_i {\rm e}^{  \a_i\cdot x }\big) = \r \cdot x   +
\ha \ln F
\   ,   $$
where the constants $\ep_i$ can be chosen to be $ 0$ or $ \pm 1$ and $\r= \ha
\sum_{s=1}^m  \a_s$ is
 half  of the sum of all positive roots.\foot{ If  $\a_1$  is  a simple root
corresponding to the generators $E_{\pm\ao}$
which are left ungauged  (the remaining  $m-1$  positive (negative) roots
correspond to the generators of a  left (right) nilpotent subgroup that was
gauged)
then $\ep_1=1$  ($m= \ha (d-n), \ n= \rank G, \ d=\dim G $).}

\def \ax {  \a_i\cdot x }

The case of  rank 2 groups  leading to four dimensional backgrounds will be
studied  in detail in Section 3.

In Section 4  we  shall  find   the general conditions on the functions $F,\p$
which are necessary for conformal invariance and check that the functions
obtained from gauged WZNW models are their solutions.  We shall also  consider
the dual version of the above \sm,
 $$  \tilde S = {1\ov \pi \a'} \int d^2 z
 \big[   \del x^i \bd x_i   + \TF(x) \del \tu \bd \tu   -  2 \del \tv \bd \tu
   \big]   +  {1\ov 4 \pi  }  \int d^2 z \sqrt {g^{(2)}}  R^{(2)} \tp (x) \   ,
  $$
$$ \tilde F  = F\inv =  \sum_{i=1}^n \ep_i {\rm e}^{ \a_i\cdot x } \ , \ \ \
\tilde \p = \p - \ha \ln F = \r\cdot x  \ , $$
which is also an exact  solution of conformal invariance conditions
and discuss an    apparent similarity  to  the Toda model (in particular,  we
shall find a  direct relation between the solutions of the classical equations
of motion).
 Some concluding remarks will be made in Section 5.

\newsec{Null gauging of WZNW models}

\subsec{\bf General scheme}

The simplest possibility to  construct  a  $D=4$ solution using a $G/H$ gauged
WZNW model
is to consider $H$ to be a subgroup of a semisimple group $G$ of a minimal
possible dimension. Since we would like also to get  a $D=4$ space-time  with
a time-like direction   the obvious candidates  for $G$ are  {\it non-compact}
 groups, e.g.
 $SL(2,R)\times SL(2,R)$ or $SO(1,3)$.
The Killing form of the first
algebra has the signature  $(-,+,+,-,+,+)$
so that one can get  the Minkowski  signature of the $D=4$  background   by
the standard  (vector or axial) gauging of one compact
and one non-compact generator.

 There exist, however, another possibility which we shall exploit below.
 The indefinite signature of the Killing form for non-compact  algebras implies
that
there is a number of `$null$'  generators  $T_n=N_n$  which have zero scalar
products,
$ \Tr (N_nN_m) = 0$. A subalgebra generated by such generators
is thus solvable (but may not be nilpotent).\foot{An example of  a `null'
 generator in the Lorentz group case
is  a  sum of a spatial rotation with a boost. Note that a nilpotent ($N^2=0$)
generator  is null but, in general, a null generator need not be nilpotent.
Gauging of subgroups generated by nilpotent generators
was previously discussed  in \oraif\dvv\jack\ira\arfa\ind.}
  In this  case  one can  consider  a  left-right asymmetric gauging
since the anomaly cancellation condition \witte\  ($\Tr T_L^2=\Tr T_R^2$)
is   obviously satisfied.

\def \N { \bar N}
Let us first  recall the structure of the action of the standard vectorially
gauged $G/H$  WZNW model.
The classical  $G/H$ gauged  WZNW action \witt\gwzw\gwz\karabali\ \foot{For our
discussion of the non-compact group gauging  it is useful to  choose the
`unphysical' sign in front of the $2d$ action,  with $k$  now being positive.
This  implies that the Cartan subgroup coordinates
will have positive signs in front of their kinetic terms.}  $$
S_v = -  k I_v (g,A)  \ , \ \ \
I_v (g,A) = I(g)  +{1\over \pi }
 \int d^2 z \Tr \bigl( - A\,\bd g g\inv +
 \bar A \,g\inv\del g  $$ $$ + g\inv A g \bar A  - A \A \bigr)
 \equiv  I_0(g,A) -
 \1p \int
d^2z \Tr ( A\A )  \ ,   \eq{1} $$
 $$    I \equiv  {1\over 2\pi }
\int d^2 z  \Tr (\del g^{-1}
\bd g )  +  {i\over  12 \pi   } \int d^3 z \Tr ( g^{-1} dg)^3   \ .
$$
is invariant under  the    vector $H$ -  gauge transformations
 $$ g \ra w^{-1} g \bu \ , \  \ A \ra w\inv ( A + \del ) w\ , \ \
 \A \ra \bu\inv ( \A + \bd ) \bu \ , \ \  \ \ \bu =w= w (z, \bar z) \  . \eq{2}
$$
 Parametrising  $A$ and $\A$ in terms of $\  h$ and $\bh$  from $H$
$$ A = h \del h\inv \  , \ \ \ \A = \bh
\bd \bh\inv  \  ,
 \ \ \ h \ra w\inv h \ , \ \ \bh \ra   \bu\inv \bh  \  ,\eq{3} $$
one can use  the  Polyakov-Wiegmann identity \polwig \ to represent the gauged
action   as the
difference of the two manifestly gauge-invariant terms: the ungauged  WZNW
actions
corresponding to the group
$G$
and  the subgroup $H$,
$$ I_v (g,A) = I (h\inv g \bh ) -  I (h\inv \bh)  \  . \eq{4} $$
Since
 $$ I(h\inv g \bh ) = I_0 (g,A)  + I(h\inv) + I(\bh)  , \ \   \eq{5} $$ $$
  I(h\inv \bh) =  I(h\inv) +
I(\bh)  + \1p \int d^2 z \Tr (A\A)   \  , $$
 the non-local terms $I(h\inv) + I(\bh)$  cancel
out in   the  classical action (4) but  survive in the quantum effective one
\ts\bs\ since
the  coefficients $k$
of the two terms in (4) get  different quantum  corrections ($k \ra k- \ha c_G$
and $k\ra k- \ha c_H  $).\foot{$c_G$ is the value of the quadratic Casimir
operator in adjoint representation.
The negative sign of the shift is due to our choice of the `unphysical' sign in
the action  (1) as usual in the non-compact case.}
The presence of the two WZNW terms in (4) with the coefficients which
renormalise in a different way  implies that the  target space background
fields obtained after  integrating out the gauge field \bcr\wiit\  are modified
by  $k\inv (\sim \a')$ -corrections \ts\bs.

Suppose now  that the parameters $w$ and $\bu$ of the gauge transformation in
(2) are not the same and belong
to two  different  subgroups $H_+$ and $H_-$ of $G$. If these subgroups are
generated by null generators ($ \Tr (N_nN_m) = 0$) we have the crucial property
 $$I(h\inv)= I(\bh)=0\ . \eq{6} $$
Then one
may use  the action
$$ S_n = -k I_n \ , \ \ \ \  \ \ I_n (g,A,\A)\equiv  I (h\inv g \bh ) $$  $$=
I(g)
 +{1\over \pi }
 \int d^2 z \Tr \bigl( - A\,\bd g g\inv +
 \bar A \,g\inv\del g + g\inv A g \bar A   \bigr) \  ,   \eq{7} $$
which is manifestly {\it gauge invariant} and {\it local}  when  expressed in
terms of $g$ and   $A$, $\A$,\foot{$A$ and $\A$  should be considered as chiral
projections of the  two independent vector fields. }
 as the  gauged WZNW action.\foot{ Depending on a choice of the null subgroups
$H_+$ and $H_-$  the trace of the product of their generators $\Tr (N\N)$  and
hence  $ \Tr (A\A)$ may or may not vanish so that (7), in general, is different
from (4) (which is not gauge invariant if $\Tr (N\N)\not=0$).
 In the special case of (6) the  action (7) coincides with the action of the
chiral gauged WZNW model \tye\sfetsos\sfts\
$\  I_c (g,A) = I (h\inv g \bh ) -  I (h\inv ) - I(\bh)$. }
Assuming that  the  corresponding quantum theory is regularised in the
`left-right decoupled' way (so that  the local
counterterm  $ \Tr (A\A)$   does not appear)
the  only  non-trivial renormalisation that can  occur at the quantum level is
the shift  of the overall coefficient
$k\ra k- \ha c_G$ in front of the action (7).  As a result, the  couplings of
the \sm
obtained by integrating out the gauge fields $A,\A$   should  not receive
non-trivial $k\inv$ -corrections,
i.e. they should represent an  exact solution of the  \sm conformal invariance
equations.
The central charge of the resulting gauged model will be equal to the central
charge of the original WZNW model minus the  dimension of the gauged subgroup.

\subsec{\bf Gauging of nilpotent subgroups  in  Gauss decomposition
parametrisation}
A  particular case of such gauging (when the null subgroups are the nilpotent
subgroups corresponding to  the step  generators in the Gauss decomposition)
was  considered previously \oraif\dvv (see also \jack\ira)  in the context of
Hamiltonian reduction \shat\ber\
of WZNW theories related to  Toda models.\foot {The WZNW model in the Gauss
decomposition parametrisation was considered in \ger. The standard (vector or
axial) gauging  in the  Gauss decomposition  was  also discussed  in \arfa.}
The  approach based on gauging of any  subgroup  with null generators
 is more general since, in principle,  we
do not need  to use  the Gauss decomposition (which does not always exist for
the real groups  we are  to consider to  get  a real WZNW action).
 The gauging based on the  Gauss decomposition   directly applies only to the
groups  with the algebras that are  the `maximally non-compact' real forms of
the classical Lie algebras
(real linear spans of the Cartan-Weyl   basis), i.e.   $sl(n+1,R), \ so(n,n+1),
\ sp(2n,R), \ so(n,n)$. The corresponding WZNW models can be considered as
natural generalisations of the $SL(2,R)$ WZNW model.\foot{Since there exists a
Cartan involution for every
non-compact real form of  complex simple Lie algebras  \helgas\ it may be
possible to repeat the construction  that follows for other non-compact groups
using `generalised Gauss decomposition'  \feher. Our treatment of the $SO(1,3)$
case below  may be considered as  a particular example  of such  a
generalisation. }  For these groups there exists a real group-valued Gauss
decomposition
$$  g= g_+  g_0 g_- \ , \ \ \  g_+ = \exp ( \sum_{\Phi_+ } u^\a E_{\a }) \ , \
\
g_-= \exp ( \sum_{\Phi_+} v^\a E_{-\a }) \ ,  \eq{8} $$
$$   g_0 = \exp ( \sum_\Delta  r^\a H_\a )=   \exp ( \sum_{i=1}^n  x^i H_i)  \
. $$
Here $\Phi_+$ and $\Delta$ are the sets of the positive and simple roots of a
complex algebra with the Cartan-Weyl  basis consisting of the  step operators
$E_\a, \ E_{-\a}  , \ \a \in \Phi_+ $ and  $n(=\rank G)$  Cartan  subalgebra
generators $ H_\a , \ \a\in \Delta$.\foot{In  what follows we shall assume
that there is always  a sum over  repeated upper and lower indices.
We shall also  use $r^\a$  with understanding   that $r^\a \not=0$ only if $\a$
is a simple root.}
We shall use the following  standard relations  (we  shall assume that a long
root has $|\a|^2 =2$)
\hum\olive\
$$
[ H_\a,E_\b]= K_{\b\a} E_\b \   \  ( \a \in \Delta, \  \b \in \Phi )\  ,
  $$ $$
  [E_\a, E_{-\a} ]= H_\a     \   \ ( \a\in \Delta  ) \    , \ \   \
[ E_\a,E_\b]= N_{\a\b} E_{\a + \b}  \ , $$
$$ \Tr (H_\a H_\b ) \equiv C_{\a\b}={2\ov |\a|^2} K_{\a \b}\ , \ \ \  K_{\a
\b}= {2 \a \cdot \b \ov |\b|^2} \ ,   \   \ \ \Tr (H_i H_j ) =\d_{ij}  \ ,   $$
$$ H_\a = \tilde\a^i H_i  \ , \ \ \
 \ , \ \ \   x^i =   \sum_{\a \in \Delta} \tilde\a^i r^\a \ , \ \  \  K_{\a\b }
r^\b  =\a\cdot x \ , \ \ \ \tilde\a^i\equiv  {2\ov |\a|^2} \a^i \ ,  $$
$$ \Tr (E_\a E_{-\b }) =  {2\ov |\a|^2} \d_{\a \b} \ , \ \   \
 \Tr (E_\a H_\b )=0 \ , \   \eq{9} $$
where  $\a^i \ (i=1,...,n)$ are the components of the positive  root vectors.
It is  clear that $E_\a$ and $E_{-\a}$ form sets of null generators  so that
some
of the corresponding symmetries can be gauged  according to (7). For example,
we   may take $w, \ A $ and $h$ in (2),(3)  to  belong  to the one-dimensional
subgroup generated by some $E_{\g}$    and  $\bu, \A $ and $\bh$ -- to the
subgroup generated by
$E_{-{\g'}} $   where $\g$ and ${\g'}$ are  positive roots which  may not
necessarily be the same.

If one gauges  the full left and right  nilpotent subgroups  $G_+$ and $G_-$
($\dim G_{\pm} = \ha (d-n)= m, \ \dim G = d ) $  generated by
all generators $E_\a$ and $E_{-\a}$ \oraif\ one is left with the action for
$n$  decoupled scalars $r^\a$  or $x^i$
 which represent  the free part of the  Toda model action.
Being interested in finding non-trivial  conformal  \sms
describing string  solutions we  are to consider the more general case of
`partial' gauging  when   only some
subgroups   $H_+$ and $H_-$  of  $G_+$ and $G_-$ are gauged.
  $r^\a$  should correspond to spatial directions ($C_{\a\b}$  in (9) is
positive definite).
Since  the Killing form  of the maximally non-compact groups has $m=\ha (d-n)$
time-like directions,  to get a physical signature  of the  resulting
space-time  we need to gauge
 away all but {\it one}  pair of coordinates $u,v$ in (8). Therefore
 the gauge groups $ H_{\pm}$  should  have dimension $m-1=\ha (d-n)-1$, i.e.
$$ \dim H_\pm = \dim G_\pm - 1  \ . $$
As we shall see below (in Sect.2.3) the ungauged generator(s) of  $G_\pm$
must be a simple root.
Moreover, to get   the physical value    $D=4$  of the target space  dimension
we need to  start with  the  rank 2 groups  $G$  ($D= n + 2=4$).

Let us first consider the most general case   when
 $w, \ A $ and $h$ in (2),(3)   correspond to the subgroup   $H_+ \subset G_+$
generated by some  $s\leq m$
 linear combinations  ${\cal E}_p = \l^\a_p  E_\a $ of the generators  $(E_\a,
\ \a\in \Phi_+ ) $  of $G_+$
and  $\bu, \A $ and $\bh$ -- to the subgroup $H_- \subset G_-$ generated by
some  $s'=s$ linear combinations $\EE_{q}= \ll^\a_q E_{-\a} $.
   Then  it is straightforward to write down the resulting  expression for the
action  (7)  using
the Polyakov-Wiegmann formula  and (9) (i.e.  $I(g_+)=I(g_-)=0$, etc.)
$$ I_n = I (h\inv g \bh) = I(g_0)  +  {1\over \pi }
 \int d^2 z \Tr \bigl[  g_0\inv  g_+\inv h \del (h\inv g_+) g_0  g_-\bh \bd
(\bh\inv g_-\inv)
\bigr]
$$ $$ = I(g_0)  + {1\over \pi }
 \int d^2 z \Tr \bigl[  g_0\inv (A +  g_+\inv \del  g_+) g_0  (\A -  \bd g_-
g_-\inv)
\bigr] \ . \eq{10} $$
Setting
$$   A=  {\cal E}_p B^p = \l^\a_p B^p E_\a\ , \ \ \ \A =   \EE_q  \B^q=
\ll^\a_q \B^qE_{-\a}
 \ ,   $$ $$ \  g_+\inv \del  g_+  \equiv J_u = U^\b_\a (u) \del u^\a  E_\b \ ,
\ \ \  \bd g_-   g_-\inv \equiv \J_v = \J^\a_v E_{-\a} = V ^\b_\a (v) \bd v^\a
E_{-\b}  \ ,  \eq{11} $$
we get
$$ S_n =  {k\ov \pi} \int d^2 z  \big[\ha C_{\a\b} \del r^\a \bd r^\b
 +   M_{\a\b} (J^\a_u  + \l^\a_p B^p) (\J^\b_v - \ll^\b_q \B^q) \big] \  $$
$$ =  {k\ov 2\pi} \int d^2 z  \big[ \del x^i \bd x_i
 +   2M_{\a\b} (J^\a_u  + \l^\a_p B^p) (\J^\b_v - \ll^\b_q \B^q) \big] \   ,
\eq{12}    $$
where
$$ M_{\a\b} = \Tr ( g_0\inv E_\a g_0 E_{-\b} ) = f_\a (r)  \d_{\a\b}  \ , \ \ \
f_\a (r)  \equiv  {2\ov |\a|^2} {\rm e}^{  - K_{\a\b } r^\b } ={2\ov |\a|^2}
{\rm e}^{  - \a\cdot x } \ .  $$
The  sums over $\a,\b$  run  over  positive roots ($r^\a\not=0$   for simple
roots only). It is clear that when  $H_\pm$=$G_\pm$, i.e. when $\l^\a_p$
and $ \ll^\b_q  $  are non-degenerate  we can eliminate $J_u$ and $\J_v$ from
the action by redefining  the gauge fields $B, \B$. One is then left with the
free action for $r^\a$
plus the dilaton term  $\p=  \p_0 +  \ha \sum_\a K_{\a\b } r^\b $
originating from the $B,\B$ -determinant. More precisely, the latter
determinant is given by the sum of the {\it two}  terms \schts\ so that its
contribution to the action is
$$ \Delta S =  - {1\ov 2\pi} \int d^2 z\  \del  ({\rm ln\ } \det M)  \bd  ({\rm
ln\ }  \det  M)
 -    {1\ov 8\pi  }  \int d^2 z  \sqrt{ g^{(2)}} R^{(2)} \ln \det  M \  $$
$$  = - {1\ov 2\pi} \int d^2 z  \sum_{\a,\b\in \Phi_+}  (\a\cdot \del x )(
\b\cdot  \bd x )
   +   {1\ov 4\pi  }  \int d^2 z  \sqrt{ g^{(2)}} R^{(2)} (\p_0  + \ha \sum_{\a
\in \Phi_+} \a \cdot x  )
$$ $$ = - {2\ov \pi} \int d^2 z    (\r \cdot \del x )( \r \cdot \bd   x )
  +  {1\ov 4\pi  }  \int d^2 z  \sqrt{ g^{(2)}} R^{(2)}  (\p_0  +  \r \cdot x
)
 \ ,  \ \ \ \   \r\equiv  \ha \sum_{\a\in \Phi_+} \a  \ . \eq{13} $$
Similar expression was found in \ger\  in the process of representing WZNW
theory in terms of free fields. It was claimed in \ger\ that the first term
combined with the free term in (12)
produces the  quantum renormalisation of the level coefficient $k$ leading to
the correct expression for the  total central charge.
As it appears from (13), only the coefficient of the projection of $x$ on $\r$
gets renormalised
(this, in fact, is  sufficient   for reproducing the quantum value of the
central charge  of the WZNW model
$C= k/(k- \ha c_G), \  c_G= 24\r^2/d)$.  In what follows we shall  not include
explicitly the derivative terms in similar gauge field  determinant
contributions   anticipating that
the quantum effective action of the resulting  model  has the  shifted overall
coefficient
$\k=k-\ha c_G$.

Integrating out $B^p$ and $\B^q$  in (12) we get
$$ S_n =  {k\ov \pi} \int d^2 z  \big[\ha C_{\a\b} \del r^\a \bd r^\b
 +   \M_{\a\b} (r)   U^\a_\g (u)  V^\b_\d  (v) \del u^\g  \bd v^\d  \big] $$ $$
  -  {1\ov 8\pi  }  \int d^2 z  \sqrt{ g^{(2)}} R^{(2)} \ln \det  M_{pq} (r)  \
  \  ,  \eq{14 }  $$
$$  M_{pq}(r) \equiv   M_{\a\b}  \l^\a_p  \ll^\b_q = \sum_\a  f_\a (r)  \l^\a_p
 \ll^\a_q \ , \ \
 \   $$ $$  \M_{\a\b} (r) =M_{\a\b} - M^{-1 pq} \ll^\g_p \l^\d_q
M_{\a\g}M_{\b\d} =
 f_\a   \d_{\a\b} -   f_\a f_\b M^{-1 pq} \ll_{\a p } \l_{\b q}   \ .  \eq{15}
$$
For example, in the simplest  case  when   the left and right  null  gauged
subgroups  are one-dimensional and  generated by the  opposite roots $E_\g,
E_{-\g}$
we get
$$   S_n = {k\ov 2\pi} \int d^2 z  \big[ \del x^i
\bd x_i  + \sum_{\a\not=\g }
{4\ov |\a|^2} {\rm e}^{  - \a\cdot x } U^\a_\s (u)  V^\a_\d  (v) \del u^\s  \bd
v^\d  \big]
 $$ $$+  {1\ov  4 \pi  }  \int d^2 z \sqrt {g^{(2)} } R^{(2)}  (\r- \ha
\g)\cdot x  \ .    \eq{16} $$
The derivatives of  the corresponding coordinates $\del u^\g $  and  $\bd v^\g$
are absent in  the action  (which is just the original ungauged WZNW action  in
(12) with
$J^\g_u=\J^\g_v=0$)
but  $u^\g, v^\g$ themselves
 may still  appear in (16)  through $U^\a_\s (u)  V^\a_\d  (v)$.
This  does not represent a problem    for a \sm interpretation  since
one  can   gauge fix $ u^\g=v^\g=0$  from the start (cf. \arfa).
The null gauging thus reduced the dimension of the WZNW model by two
as expected since the left and right gauge groups are independent.\foot{ In
general,  $H_+$ and $H_-$ need not be the same so that $M_{pq}$
may be degenerate. The integrals over the corresponding
 zero eigenvalue combinations of $B,\B$ will produce $\d$-function
constraints on some of $J_u$, $\J_v$.  In the case when the gauge groups are
one-dimensional and the  corresponding left and right gauge generators  are
just different
 step operators $E_\g$ and $E_{-\g'}$,
 the $B\B$-term  in (12) vanishes  and the integrals over $B$ and $\B$ give the
$\d$-function  constraints which  set  $\J_v^\g $ and $J_u^{{\g'}}$ to zero
($M_{\a\b}$ is diagonal) so that the final result for a \sm action is
$   S_n = {k\ov 2\pi} \int d^2 z  \big[  C_{\a\b} \del r^\a \bd r^\b   +
\sum_{\a\not=\g , {\g'} }   f_{\a } (r) U^\a_\s\ (u)  V^\a_\d  (v) \del u^\s
\bd v^\d  \big]    \ .  $
There is an extra  dilaton term originating
from the  determinants which  appear after integrating  out the $\d$-functions.
 This `degenerate' gauging reduces the number of dimensions by $four$:  $\del
u^\g,  \del u^{\g'}, \bd v^\g, \bd v^{\g'}$ are absent in  the final  action
(14).  In fact, the  built-in gauge invariance of the original action
(10) implies that $u^\g$ and $v^{\g'}$  can be gauge-fixed to zero.
 Also,  given that $\del u^{\g'}$ and $\bd v^\g$ appear in the final result
only in the $\d$-functions, it is natural to   reduce the number of coordinates
further by  trying   to  integrate explicitly
over $u^{\g'}$, $v^\g$.}

If the gauge fields in  (12) belong to $\Phi/\Delta$
then after integrating them out we finish with
the interaction term $2M_{\a\b}J^\a_u \J^\b_v$
where $\a, \b$ run over simple roots only. If one further adds the constraints
\oraif\
on these currents  (which  break manifest  off-shell
classical  conformal invariance  of the  gauged WZNW model) one finds that
(12) takes the form of  the Toda model. As we shall see in Section 4.2 (see
also Section 2.4), the  imposition of constraints is not, in fact, necessary in
order to make a connection to the Toda model.

In the case  when gauged subgroups are non-abelian
the resulting \sms\  (14)   may have non-abelian global symmetries and
in particular cases have classical equations equivalent to equations in
non-abelian Toda models (see, e.g.,  \gervais\wip\feher\natd\jackk). However,
such  models will contain more than one pair of $u,v$ coordinates, i.e. more
than one  time-like direction and we shall not consider them here.

\subsec{\bf Models with one time-like coordinate}

Let us now  turn  to the most interesting case
when  the dimensions of the gauge groups $H_\pm$ are equal to $\dim G_\pm
-1=m-1$
so that only one time-like  coordinate appears in the  resulting \sm action.
Let  $E_\ao$  and $E_\am $  denote the generators of $G_+$  and $G_-$  which
remain ungauged, i.e.
which do not belong to $H_+$ and $H_-$. Since $H_+$  must  be a subgroup,
 $E_\ao$  cannot appear in the commutators of the generators
of $H_+$.  According to (9) this is possible only if  $\ao$ is a simple root,
i.e.  if it cannot be represented as a sum of two other positive roots.
In fact, if  we use the indices $i,j$ to denote the simple roots $\a=\a_i \
(i=(1,s)=1,2,...,n)$   and indices $a,b$ to denote the remaining positive roots
$\a_a \ (a=n+1,...,m)$  the commutators of the corresponding  step operators
are given by
$$ [E_i,E_j] \sim E_a \  \  (\a_a= \a_i + \a_j) \  ,  $$ $$  [E_i, E_a] \sim
E_b \ \   (\a_b= \a_i + \a_a)\ , \  \  \  [E_a, E_b] \sim E_c \  \  (\a_c= \a_a
+ \a_b) \   . $$
It is  clear that one can also use  linear combinations $E_s'= E_s + \l_s E_1 \
(s=2,...,n)$
as the  `simple' part of the generators of $H_+$ but one cannot mix the
non-simple generators $E_a$ with $E_1$.

Let $ u^\ao \equiv  {1\ov \sqrt 2}u, \ v^\ao \equiv {1\ov \sqrt 2}  v$; the
remaining coordinates $u^\s, v^\s$ ($\s=(s,a)$  will be used to denote  all
`gauged'  $m-1$  positive roots)
 are transforming under the gauge group (with the leading-order term being just
a shift)
so that we can set them to zero as a gauge. In this gauge $J_u =  {1\ov \sqrt
2}\del u E_\ao, \
\J_v =  {1\ov \sqrt 2} \bd  v E_\am$ and the \sm  action  (14) takes the  form
($p,q=1,...,m-1$)\foot{For notational convenience (to get rid of an extra
factor of 2 in front of the $  F (x)   \del u  \bd v  $ term)  we have
redefined $u$ and $v$ by the factor of $1/\sqrt 2$ as compared to (14). }
$$ S_n =  {k\ov 2\pi} \int d^2 z  \big[ \del x^i \bd x_i
 +   F (x)   \del u  \bd v  \big]    +  {1\ov 4\pi  }  \int d^2 z  \sqrt{
g^{(2)}} R^{(2)}  \p (x)  \ ,   \eq{17 }  $$
$$  F(x) = f_\ao   -   f^2_\ao  M^{-1 pq} \ll^\ao_p \l^\ao_q  \ , \ \ \p (x) =
- \ha \ln \det  M_{pq} \ , $$ $$ \ \ M_{pq}(x) = \sum_\a  f_\a (x)  \l^\a_p
\ll^\a_q \ . $$
The non-trivial elements of the `mixing' matrix  $ \l^\a_p  $
correspond to a possibility of changing the generators of $H_+$ by adding
$\l^\ao_p E_\ao $.
Without loss   of  generality   the non-vanishing components of $\l^\a_p$ can
be taken  to be:
 $ \l^\s_p = \d^\s_p , \ \  \l^\ao_p = \l_s \d_{ps}$     and similarly for
$\ll^\a_q$ (according to the remark above, only simple roots can be mixed with
$E_\ao$).
 Then  ($s,t=2,...,n; \ a,b= n+1,...,m$)
$$  M_{pq}=    \left(\matrix
{M_{st} &0\cr 0&M_{ab}\cr } \right) , \ \
 M_{st} (r) = f_s \d_{st}   +   f_1  \l_s  \ll_t  \ , \ \ M_{ab} (x) =  f_a (x)
\d_{ab}\ , \  \eq{18}  $$
$$
f_h (x) \equiv f_{\a_h} = {2\ov |\a_h|^2} {\rm e}^{  - \a_h \cdot x   }  \ , \
\ \   \ h=(1,s,a)=1,2,...,m \ , \ \ m
=\ha (d-n)\  .
 \eq{19} $$
If we  introduce  $\l_1=\ll_1 =1 $  in order to make the formulas look
symmetric with respect to all simple roots,
we find
$$ M_{st}\inv =  f_s\inv \d_{st}   -
    { f_s\inv f_t\inv \l_s  \ll_t \ov  \sum_{i =1}^n f\inv_{i} \l_i \ll_i} \ ,
\ \ \  \det M_{pq} =  \big(\prod_{h=1}^m f_h \big) \big(\sum_{i =1}^n f\inv_i
\l_i \ll_i  \big)\ . \eq{20}   $$
As a result,
$$    F  = f_1   -   f_1^2  M^{-1 st} \ll_s \l_t  = { 1 \ov
 \sum_{i =1}^n f\inv_i \l_i \ll_i }  \ , $$ $$  F  =
 \bigl( \sum_{i =1}^n  \ep_i  {\rm e}^{  \a_i \cdot x  } \bigr)\inv \ ,  \ \ \
\  \  \ \ep_i\equiv
 \ha {|\a_i|^2} \l_i \ll_i \ ,   \eq{21} $$
$$
 \p= - \ha \sum_{h=1}^m \ln  f_h  - \ha  \ln  \sum_{i =1}^n f\inv_i \l_i \ll_i
\ ,  $$ $$
\p =\p_0   +  \r \cdot x   -
\ha  \ln   \sum_{i =1}^n  \ep_i  {\rm e}^{  \ax  }  \ , \ \ \ \  \r= \ha
\sum_{h=1}^m  \a_h \ .
    \eq{22} $$
We    discover the following relation  between the basic function $F$ and the
dilaton
$$ F (x) =   F_0  {\rm e }^{2\p (x) }   {\rm e }^{ - 2\r\cdot x }
\ . \eq{23} $$
This relation is not accidental, being  necessary for the
  conformal invariance of the model (17) (see  Section 4). Note  that (23)
implies that  the  string  tree-level measure  factor for the model (17)
 has a universal  form which does not depend on a particular gauging but only
on
the sum of all positive roots for a given algebra
$$ \sqrt G   {\rm e }^{- 2\p } = F {\rm e }^{- 2\p } = F_0   {\rm e }^{ -
2\r\cdot x }  \ . \eq{24} $$
This factor, in fact, is the $x$-dependent part of the Haar measure  of the
original WZNW model  expressed in the Gauss decomposition (see (14),(15)).

It is clear that non-equivalent models correspond   only to  $\ep_s=0, +1, -1$
since if $\ep_s\not=0$ we can make  all $|\ep_s|=1$ by constant
shifts of $x^i$.
In the simplest case when all  mixing parameters $\l_s, \ll_s$ are  equal to
zero, i.e. $\ep_s=0$,
$ F=f_1  , \  \p(r) = \p_0 - \ha \sum_{s=2}^m  \ln f_h   \ ,  $
 the action (17) becomes (cf.(16))
$$ S_n =  {k\ov 2\pi} \int d^2 z  \big[ \del x^i \bd x_i
 +    F_0 {\rm e}^{  - \a_{1i} x^i  }  \del u  \bd v  \big]    +  {1\ov 4\pi  }
 \int d^2 z  \sqrt{ g^{(2)}} R^{(2)}   (\r_i - \ha \a_{1i}) x^i     \ ,
\eq{25 }  $$
where $\a_{1i}$ are components of the  `ungauged'  simple root.
Using the rotational symmetry in $x^i$ space  we can make $\a_1= (\sqrt 2, 0,
...,0)$  so that  the model factorises into a product of the $SL(2,R)$ WZNW
model
and $n-1$ free scalars with linear dilaton.
\subsec{\bf Summary}

We   have  thus found   the \sm action  (17),(21),(22), i.e.\foot{We absorb the
prefactors $
2/|\a_i|^2 $ of  the exponential  terms into a rescaling of $u,v$ and $\ep_i$.
We also consider (26) as an effective action, including the  quantum shift of
$k$,
 $ \ \k\equiv k- \ha c_G$. }
$$ S_n =  {\k\ov 2\pi} \int d^2 z  \big[ \del x^i \bd x_i
 +      \bigl( \sum_{i =1}^n  \ep_i  {\rm e}^{  \a_i\cdot x  } \bigr)\inv  \del
u  \bd v  \big]    $$
$$ +  {1\ov 4\pi  }  \int d^2 z  \sqrt{ g^{(2)}} R^{(2)}  \big( \r\cdot x    -
\ha
  \ln   \sum_{i =1}^n  \ep_i  {\rm e}^{  \a_i\cdot x   } \big)   \ ,    \eq{26
}  $$
with $\a_i$ being the simple roots and $\r$ being  half the sum of all positive
roots.  The  values of the  parameters
$\ep_i=0, +1$ or $ -1$
  represent  inequivalent   gaugings of the original WZNW model or different
conformal sigma models.\foot{
Inequivalent solutions  corresponding to different possible choices of
an   ungauged simple root $\a_1$  are easily included by assuming that $\ep_1$
can also take values $0$ and $-1$ but at least one of $\ep_i$ is non-vanishing.
 In general,   $\ep_i$  taking arbitrary real values  represent  moduli of the
solutions.}  Non-trivial models (not equivalent to direct products of $SL(2,R)
$ WZNW with free scalars) are found for non-vanishing values of the  `mixing'
parameters $\ep_2, ...,\ep_n$.

 The  metric of the corresponding $D=n+2$ dimensional target space-time has two
null Killing  symmetries (in fact, the full $2d$ Poincare  invariance in the
$u,v$ plane, $\ u'= \r u + a, \ v' = \r\inv v + b$).
The    non-trivial $(uv)$  components of the metric and  the antisymmetric
tensor
and the non-linear part of the dilaton  are all  expressed in terms of a
single function   $F(x)$  (21),
which is the inverse of the sum of  the exponentials of the spatial Cartan
coordinates $x^i$. The metric is non-singular if all $\ep_i$ have the same
sign.

 The action (26) has a structure  reminiscent  of the Toda model (with  the
dimension 2  operator
  $\del u  \bd v$  instead of  the dimension zero one  in the  interaction
term).
In general, the model (26) or its dual version which has  the metric and
dilaton given by (see  Section 4.3)
$$\tilde F =  F\inv= \sum_{i =1}^n  \ep_i  {\rm e}^{ \ax }\ , \ \ \ \
\tilde \p = \p - \ha \ln F = \r \cdot x  \ , \eq{27} $$
 can be considered as  `\sm  analogs' of the  Toda model.

We are drawing
an analogy  with abelian Toda models (see, e.g., \lez\mans\oltu\bil\oraif).
In non-abelian Toda models (see, e.g.,  \gervais\wip\feher\jackk\natd)
the free Cartan subgroup $x^i$-part of the  abelian Toda action is  replaced by
a WZNW model action (and, correspondingly, the potential term takes more
complicated form).
 As noted at the end of Section 2.3, our discussion can be generalised to the
case  of non-abelian gauge subgroups  so that the resulting \sms  (in the
special cases of properly chosen subgroups
as in  the generalised integral gradings gauging in
\wip\oraif\feher)
 have  a structure  similar to that of   non-abelian Toda models (being
equivalent to them
at the level of classical equations of motion).
However, such  models  have more than one time-like direction and  do not  lead
to
new
examples of  $D=4$ conformal invariant
 backgrounds (any non-trivial WZNW
model
has  at least three degrees of freedom while a four dimensional
space has  only two transverse directions).  In fact, one can show that the
simplest model of this type  obtained by starting with $SL(3,R)$ WZNW model and
gauging  one linear combination of  roots
on the left and on the right gives a $D=6$ \sm with the signature
$(+,+,+,+,-,-)$
(for a particular choice of the gauged generator it has global $SL(2,R)$
symmetry).

 The reason for  a connection to  the Toda model   can be understood, for
example,  by comparing our approach   to that of \oraif\ where
additional `background' terms linear in the gauge fields  (implying constraints
on the currents) were  introduced into the action  (10). They
produced  a potential term after the gauge fields were integrated out. We
instead   did not gauge the full $m$-dimensional nilpotent subgroups $G_\pm$
and as a result  got   a \sm - type interaction term for the ungauged root
directions.

As we shall show in Section 4.2, the classical equations (on a flat $2d$
background)
 for the model (26) (or its dual) reduce, in fact,   to the Toda model
equations.
The  relation between the models  is not, however, quite  precise  at the
quantum level.
 Note that in contrast to the  Toda  model
(and  the dual model (27))     the dilaton  in  (26) is, in general,  a
non-linear function  of $x^i$.
If  we  gauge  the remaining  nilpotent generator corresponding to $\ao$
or simply integrate over $u,v$ in (26)  we   cancel the non-linear dilaton
term\foot{As in  (13),
the determinant  resulting from  the  integration over $u,v$ gives the dilaton
contribution  \busc\schts\  $  \Delta \p = -\ha \ln F $ which cancels
(according to (23)) the non-linear term in $\p$.}  and get  $\p= \r\cdot x$.
Though this   expression  may look similar to   the dilaton of the Toda model
in the simply-laced algebra case  \mans\bil\oraif\tsey\jack\ it is actually
different
since it does not
 contain the second linear term present in the Toda model dilaton  in the
general  non-simply-laced  case \wip\jack.
 The  reason for this disagreement  can be traced back to the basic fact that
the
interaction term in the action (26) has   classical dimension  zero while the
potential term
of the Toda model has   classical dimension  two. We shall clarify this point
in
 Section 4.2.

 In Section  4  we shall also  check explicitly  that the  background
corresponding to (26)
satisfies the conformal invariance equations.
In  the next section we shall consider the particular case of the rank $n=2$
groups  when the resulting target space has dimension  four.

\newsec{Null gauging in the case of   rank 2 groups: four dimensional
space-times}

 Let us now illustrate the above discussion on the examples of null gauging of
the groups corresponding to the maximally non-compact real forms of the rank $
n=2 $  algebras
  $sl(3,R), so(2,3)=sp(4,R)$, $so(2,2)=sl(2,R)\oplus sl(2,R) $   and $G_2$
which lead to $D=4$ backgrounds.
In the rank 2 case the action  (26)  is parametrised by  a $2\times 2$ Cartan
matrix $K_{ij}$ or by  two simple roots with components $\a_{1i}$ and $\a_{2i}$
 and  one  parameter $\ep =\ep_2/\ep_1$ with values $ \pm 1$ (we   assume that
$\ep_1\not=0$
and also $\ep\not=0$ to get a non-trivial, i.e., not a direct product
$SL(2,R)\times R^2$ model). It
has the following explicit form ($\a'= 2/\k; \ x^i=(x,y))$
$$  S = {1\ov \pi \a'} \int d^2 z
 \big[   \del x \bd x + \del y \bd y
  +  F(x,y)   \del u \bd v  \big]   +  {1\ov 4 \pi  }  \int d^2 z \sqrt
{g^{(2)}}  R^{(2)} \p (x,y)   \ ,    \eq{28} $$
$$ F=  {1 \ov   {\rm e}^{  \a_1\cdot x  }   + \ep  {\rm e}^{  \a_2\cdot x  }  }
\ , \ \ \   \  \p = \r\cdot x    -
 \ha  \ln   \bigl(  {\rm e}^{  \a_1\cdot x  }   + \ep  {\rm e}^{  \a_2\cdot x
} \bigr)   , \ \ \ \ \
\r = \ha (\a_1 +  ... + \a_m)   \ .    \eq{29}  $$
I addition to the Poincare symmetry in the $u,v$ plane this model
is also invariant under a correlated constant shift of $x^i$
and $u$ (or $v$).
Below we shall consider the particular cases of (28),(29) for all inequivalent
choices of the Cartan matrices. In the last subsection we shall discuss the
backgrounds obtained by null gauging of  the non-compact group $SO(1,3)$
which does not  admit the  standard Gauss decomposition.

\subsec{\bf  $SO(2,2)\simeq SL(2,R)\times SL(2,R)$}

Let us  start with  the simplest case of  $SO(2,2)$ or $SL(2,R)\times SL(2,R)$
when it is easy to repeat the above analysis explicitly from the very
beginning.
For  $G=SL(2,R)$\foot {The whole group
 $SL(2,R)$  can be covered by four patches, i.e. a generic element is given by
$g=  {\rm e}^{uE_+}  {\rm e}^{r H} {\rm e}^{vE_-} \omega\ , \   \   \omega =
\pm   1 \ , \ \  \omega   =
\pm \left(\matrix
{0&-1\cr 1&0\cr } \right)\ .  $   Note that  for $SL(2,R)$
  $C_{\a\b}= K_{\a\b}=2, \  |\a|^2=2, \  f_\a = {\rm e}^{-2r} $.}
 $$ g = {\rm e}^{uE_+}  {\rm e}^{r H} {\rm e}^{vE_-} =
 \left(\matrix{1&u\cr 0&1\cr }\right)
\left(\matrix{{\rm e}^{r}  &  0\cr    0  & {\rm e}^{-r}    \cr }\right)
\left(\matrix{ 1 &  0\cr  v  & 1 \cr }\right) , \ \eq{30}    $$
 $$  S=  {\k\ov \pi} \int d^2 z  \big( \del r \bd r
  + {\rm e}^{-2r}  \del u \bd v  \big)    \ .    \eq{31} $$
Gauging  the null generators of translations of $u$ and/or  $v$ we get
(the free part of) one-dimensional
 Liouville model \oraif\dvv (see also \ira).   Gauging  independently
the  left and right null subgroups  of $SL(2,R)\times SL(2,R)$  or  of
$SO(2,2)$
in a `twisted' way
we get the  following action  ($r_1,u_1,v_1$ and $r_2,u_2,v_2$ are the
parameters of the two $SL(2,R)$ groups;  $ q= \k'/\k$ is the ratio of the
corresponding $\k=k-2$ - factors for the two $SL(2,R)$ groups which may be
considered as a free parameter of the  Cartan matrix in this non-simple case)
$$ S_n =  {\k\ov \pi} \int d^2 z  \big[ \del r_1 \bd r_1
  + {\rm e}^{-2r_1}  (\del u_1 + \l B) (\bd v_1 -  \ll \B)  $$ $$
  +  q\del r_2 \bd r_2
  +  q{\rm e}^{-2r_2}  (\del u_2 +  B) (\bd v_2  -  \B)
  \big]    \ .    \eq{32} $$
This is the direct analog of (12),(13) in the $SL(2,R)\times SL(2,R)$
 case with $\l, \ll$ being the parameters of the `mixing' of the root operators
in the generators of the left and right gauge subgroups.
The action is  invariant under $u_1'=u_1 + \l a,  \ v_1'= v_1 + \ll b  , \
   u_2'=u_2+ a , \     v_2'= v_2 +  b    ,   \   B'=B-  \del a   ,  \  \B'=\B +
\bd b  . $
Gauge fixing $u_2=v_2=0$ and integrating over $B, \ \B$ we finish with the  \sm
action  with
$$  F =  {  q{\rm e}^{ -2r_1-2r_2} \ov   q{\rm e}^{ -2r_2} + \l\ll {\rm
e}^{-2r_1} } =
 {  1 \ov   {\rm e}^{ 2r_1} + \ep  {\rm e}^{2r_2} }\ ,  \  \ \ \ \ \ep = q\l\ll
\ ,   $$  $$
\ \  \p= - \ha \ln (   q{\rm e}^{ -2r_2} + \l\ll{\rm e}^{-2r_1}) =  \p_0 + r_1
+r_2 + \ha \ln F
\ .     \eq{33} $$
To  have the physical signature we  are to assume $ q>0$.\foot{  The model with
$q=-1$ (with signature $(+,-,+,-)$) can be transformed  into   another
conformal \sm  with one time-like coordinate  ($r_2$) by making the analytic
continuation in $t=\ha(u-v )$ to replace the $F\del  u\bd v$ part of the action
 by $F\del (z + i\tau) \bd  ( z-i\tau)$.
Related models  can be  obtained  by null gauging
of the $SO(1,3)$ group discussed below. The resulting metric is real but the
antisymmetric tensor is, however, pure imaginary (so that the classical string
equations and their solutions become complex). Because of the analytic
continuation involved,   this model may not also  have a  direct interpretation
in terms of the $SO(2,2)/H$ or $SO(1,3)/H$ coset conformal theories.
}

  When $\l$ or $\ll$ is zero we get the direct product of the
free  scalar  with the $SL(2,R)$ WZNW model.
For $\l\ll\not=0$  coefficient $\ep=q\inv\l\ll$ can be  made equal to $\pm 1$
by a shift   $r_2\ra r_2'$
so we are left with only two non-trivial possibilities for the \sm action (28)
corresponding to $\l\ll > 0 $  or
$\l\ll <0$  ($x=\sqrt 2 r_1, \ y=q^{1/2} \sqrt 2 r_2' $)
$$ F =  {  1 \ov   {\rm e}^{\sqrt 2 x}  + \ep  {\rm e}^{\sqrt 2\m y }  }  \ ,
\ \ \ \ \ \  \m\equiv  q^{-1/2} \ ,  \ \ \ep= \pm 1 \ ,    \eq{34}
$$ $$
  \p (x,y) =\p_0 +   {1\ov \sqrt 2 }x +  {1\ov \sqrt 2}\m y
- \ha \ln (  {\rm e}^{ \sqrt 2 x}
+ \ep   {\rm e}^{\sqrt 2\m y} )
\  \ .   $$
As we shall  explicitly check below, this background satisfies the conformal
invariance conditions.

\subsec{\bf $SL(3, R), \ SO(2,3), \ G_2$}

Let us now  consider the remaining cases of rank 2  maximally non-compact
simple Lie groups. For $SL(3,R)$  we have $d=8, \ n=2, $ the number of positive
roots
$m=3$  and   all roots have equal length ($ |\a_i|^2= 2$), i.e.\foot{We use the
notation
$K_{ij}\equiv K_{\a_i\a_j}$ where $\a_i$ are simple roots.}
$$  C_{ij}= K_{ij} =   \left(\matrix{2&-1\cr -1&2\cr }\right) \ , \  $$ $$  \
\a_1=(\sqrt 2 , 0), \ \ \ \a_2=(- {1\ov \sqrt 2} , {3\ov \sqrt 2}) , \ \  \
\r=\a_1 +\a_2 = ({1\ov \sqrt 2} , {3\ov \sqrt 2})  \ . \
\eq{35} $$
Let the  ungauged null generator which should  correspond to  a  simple root
have  the
index  1, another simple root -- index  2  and their sum -- index 3.
Then the functions $F, \ \p $ in (29),  become
$$  F  =  {1\ov
  {\rm e}^{\sqrt 2 x     }   + \ep    {\rm e}^{ - {1\ov \sqrt 2} x+ {3\ov \sqrt
2}y }  } \ , \
 \  \ \p =\p_0 + {1\ov \sqrt 2} x +  {3\ov \sqrt 2}y
 - \ha \ln (  {\rm e}^{\sqrt 2 x     }   + \ep    {\rm e}^{ - {1\ov \sqrt 2} x+
{3\ov \sqrt 2}y } )  \  . \eq{36} $$
As in the $SO(2,2)$  case considered above  $\ep$  can be  set equal to $\pm 1
$ by shifting
$y$ so that we get only two possible  non-trivial models.

In the case of $SO(2,3)\simeq Sp(4,R)$ or the algebra $B_2= C_2= so(5)$   there
are $m=4$ positive roots,
two of which  are simple and have  lengths 1 and $\sqrt 2$ (the other  two
positive roots also have lengths 1 and $\sqrt 2$).  The `kinetic'  matrix
$C_{ij}$ and the Cartan matrix  $K_{ij}$ are given by
(we assume that the first  simple root is the short one)
$$    C_{ij}=   \left(\matrix{4&-2\cr -2&2\cr }\right) \   ,
\ \  \
 K_{ij} =   \left(\matrix{2&-1\cr -2&2\cr }\right) \ , \  $$ $$   \ \
\a_1=(1,0)\ , \ \ \
\a_2=(-1,1) \  ,
\ \ \ \r=2\a_1 + {3\ov 2} \a_2 = ( \ha ,{3\ov 2} ) \   . \eq{37} $$
If the ungauged direction corresponds to the first simple root we get the
following expressions for the functions $F$ and $\p$  in (29)
$$ F   =
 {1\ov {\rm e}^{x}    + \ep  {\rm e}^{ -x + y } } \ , \ \ \
 \p = \p_0  + \ha x + {3\ov 2} y   - \ha \ln (  {\rm e}^{x}    + \ep  {\rm e}^{
-x + y })    \  .  \eq{38} $$
The case when  the ungauged root
is the long one  corresponds to interchanging the two exponentials in the sums
and is essentially equivalent to (38) (in the case when $\ep=-1$ we can also
change the sign of $u$ or $v$).

 For  the maximally non-compact form of $G_2$  ($d=14,\  n=2, \ m=6$) in
addition to the short ($|\a_1|^2= 2/3$) and long
($|\a_2|^2= 2$) simple roots there are  4 other positive roots (2 short and 2
long ones) and
$$    C_{ij}=   \left(\matrix{6&-3\cr -3&2\cr }\right) \   ,
\ \ \
 K_{ij} =   \left(\matrix{2&-1\cr -3&2\cr }\right) \ ,  $$ $$
 \a_1=({\sqrt{2\ov  3}}, 0)\ , \ \ \a_2=(- {\sqrt{3\ov  2}}, {1\ov
\sqrt{2}}) \ ,  \ \ \r =5\a_1 + 3\a_2=
( { {1\ov \sqrt 6}}, {3\ov \sqrt 2})  \  ,
\eq{39} $$
so that we find  (assuming that the short simple root is ungauged)
$$ F =
 {1\ov {\rm e}^{ {\sqrt{2\ov 3}}x     }    + \ep  {\rm e}^{ -
{\sqrt{3\ov 2}}x + {1\ov {\sqrt 2}} y  } }\ , \  $$ $$
 \p = \p_0 + { {1\ov  \sqrt 6}}x  +  {3\ov \sqrt 2}y
  - \ha \ln ( {\rm e}^{ {\sqrt{2\ov
 3}} x }    + \ep  {\rm e}^{ -
{\sqrt{3\ov 2}}x + {1\ov \sqrt{2}} y })
\ . \eq{40} $$
The choice of the long root as an ungauged one gives equivalent model.
The models of this subsection do not seem to admit
other  analytic continuations with real metric of Minkowski signature.

\subsec{\bf  Null gauging of $SO(1,3)$}

As we have already mentioned above, the  gauging  based on the Gauss
decomposition does not, however, exhaust the most general possible
gauging  of  subgroups with null generators. To implement the prescription
based on (7) one does not really need  to use  the Gauss decomposition  which
does not exist (at least in its standard version) for  non-maximally
non-compact real groups.
 To illustrate the procedure of the null  gauging  in this case let us
consider the null gauging of  $SO(1,3)$.
 The Killing form of the
corresponding algebra has the signature     $(-,-,-,+,+,+)$  implying that
there exist null generators.
Being restricted to the two-dimensional Cartan  subalgebra  the Killing form
has the signature $(-,+)$ so  that after gauging one left and one right null
generator one should expect to find a four dimensional target space of the
signature $(-,-,+,+)$. This background will  turn out to be
related by an analytic continuation to the one obtained  above by gauging
$SO(2,2)$ and will also have another interesting analytic continuation  of the
signature  $(-,+,+,+)$.

Starting with the Weyl basis ${(e_{mn})}^k_l = \eta_{ml}\d^k_n
-\eta_{nl}\d^k_m$  it  is possible to represent the six generators of $so(1,3)$
as $4\times 4$ real matrices
$H_i, E_i, E_{-i} \ (i=1,2)$
with the following commutation relations and traces\foot{Alternatively, one may
start with the
$sl(2,C)$ algebra in the $\s$-matrix basis and construct its real form in terms
of $4\times 4$ matrices by replacing $1$ and $i$ by  $2\times 2$ matrices
$\d_{ij}$ and   $\epsilon_{ij}$.}
$$[E_1, E_{-1}  ] =-[E_{2}, E_{-2}  ]=\ha H_1 , \ \
[E_2, E_{-1}  ] =[E_{1}, E_{-2}  ]=\ha H_2 \  , \eq{41} $$
$$[E_1, E_{2}  ] =[E_{-1}, E_{-2}  ]=0 \ , \ \ \   [H_1, E_{\pm i}  ] = \pm
E_{\pm i}\ , \ \
[H_2, E_{\pm i}  ] = \pm \epsilon_{ij} E_{\pm j}\ ,  $$ $$   [H_i, H_j]=0 \ ,
\ \ \Tr H_1^2=-\Tr H_2^2=1 \ ,     $$
$$
  \Tr E_{1} E_{-1}=- \Tr E_{2} E_{-2}=  \ha\ ,  \
\ \Tr H_iE_{\pm j} = \Tr E_iE_j = \Tr E_{-i} E_{-j} =0 \ . \eq{42}    $$
Note that (41),(42)  are   different from  a Cartan basis  relations.
 Still,  the  group element  of  $SO(1,3)$ can be parametrised in a way  that
mimics the Gauss decomposition for $SL(2,C)$  or $SO(2,2)$
$$ g= g_+ g_0 g_- = \exp (u_1 E_{1}  + u_2 E_{2} )  \exp (x_1 H_{1}  + x_2
H_{2} ) \exp (v_1 E_{-1}  + v_2 E_{-2} ) \ .  \eq{43}  $$
We shall gauge the following   left and right   transformations generated by a
pair of null generators\foot{Gauging in the case when  the left (right) groups
are generated by linear combinations of $E_i$ ($E_{-i})$ leads to equivalent
results
obtained by shifting  $x_2$  in (47) by an arbitary constant.}
$$ g' = \ {\rm e}^{  a_+  E_{1} } \  g \ {\rm e}^{ a_- E_{-1}} \ . \eq{44} $$
The gauge-invariant  action is  given by (7), (10). Using  (42),(43) we find in
this particular case
 $$ S_n=  {\k\ov 2 \pi} \int d^2 z  \big(  \del x_1 \bd x_1 -
\del x_2 \bd x_2  )  $$ $$  + {\k\over \pi }
\int d^2 z \Tr \bigl[  g_0\inv (BE_1 +  \del u_1 E_1 + \del u_2 E_2) g_0  (\B
E_{-1}
-   \bd  v_1 E_{-1} -  \bd  v_2 E_{-2})
\bigr] \ , \eq{45} $$
where we have used that   the  gauge fields $A, \A$  have the form  $B E_1, \B
E_{-1}$.
Evaluating the trace  using  (41),(42)   and fixing the gauge $u_1=v_1=0$ we
finish with
$$ S_n= {\k\ov 2\pi} \int d^2 z  \big[  \del x_1 \bd x_1 -
\del x_2 \bd x_2
$$ $$ + {\rm e}^{-x_1} \big( \cos x_2   B\B  +  \sin x_2  \del u_2 \B
-  \sin x_2   \bd v_2  B
+  \cos x_2 \del u_2 \bd v_2  \big)\big]
   .  \eq{46}  $$
After integrating out the gauge fields  the action takes the \sm form
($u\equiv u_2, \ v\equiv v_2$)
$$ S_n=  {\k\ov 2\pi} \int d^2 z  \big(  \del x_1 \bd x_1 -
\del x_2 \bd x_2  +  { {\rm e}^{-x_1}  \ov \cos x_2 } \del u \bd v \big)
$$ $$   -   {1\ov  8  \pi  }  \int d^2 z  \sqrt{ g^{(2)}} R^{(2)} \ln ({\rm
e}^{-x_1} \cos x_2 ) \ .
    \eq{47} $$
 Here  $x_1$ and $x_2$  can be shifted by arbitrary constants (e.g.
 $\cos x_2$ can be replaced by $\sin x_2$).
The corresponding background represents an exact  string solution with the
signature $(-,-,+, +)$ The target space metric in (47)  admits a
straightforward analytic continuation which has a  physical signature. If  we
set  $\ x_2 = ix_2' $  we get a real action of the same structure as (28)
with $F= { {\rm e}^{-x_1} ( \cosh x_2')\inv  }$. Furthermore, if we   change
the coordinates to
$x= {1\ov \sqrt 2}  (x_1 - x_2') , \ y= {1\ov \sqrt 2} (x_1 + x_2') $ the model
becomes equivalent to  a particular version
($k=k', \ \mu=1$, $\ \ep =1$) of the model  (34) obtained above by  gauging
$SL(2,R)\times SL(2,R).$\foot{An extra factor of 2 in $F$ can be absorbed into
$u,v$; one should also rescale the overall coefficient $k$ by 2 due to
different normalisations of the traces. The model  (34) with $\ep=-1$
can be also  reproduced  by starting with (47) and making appropriate complex
shifts of $x_1$ and $x_2$. }

Two other interesting analytic continuations of this model
are found by treating  $x_1$ or $ x_2$ as a time-like coordinate
and rotating $u\pm v$.  With
$ x_2=t, \ x_1=x, \ u= y-iz, \  v= y+ iz $ the  \sm action  (47) corresponds to
a time-dependent  background with the metric,
antisymmetric tensor and dilaton given by
$$  ds^2 = -dt^2  + dx^2  +   { {\rm e}^{-x}  \ov \cos t }  (dy^2 + dz^2)
\ , \ \ \  B_{yz} =  i{ {\rm e}^{-x} \ov \cos t }
\ , \ \ \   \p = \ha x - \ha \ln \cos t  \ . \eq{48} $$
If instead $ x_2=x, \ x_1=t, \ u= -y+iz, \  v= y+iz $  and $\k \ra -\k$ we get
$$  ds^2 = -dt^2  + dx^2  +   { {\rm e}^{-t}  \ov \cos x }  (dy^2 + dz^2)
\ , \ \ \  B_{yz} = i { {\rm e}^{-t}  \ov \cos x }
\ , \ \ \   \p =  \ha  t  - \ha  \ln  \cos x  \ .  \eq{49}  $$
Up to a  potential problem  of   imaginary nature of the antisymmetric tensor
field  (the $i$ factor in $B_{mn}$ drops out from the string  effective
equations  but may  not allow a physical  interpretation of the solution since
the string action  and the string equations  are  complex)
  these backgrounds  represent exact string solutions which  may have a
`cosmological' interpretation.\foot{While completing this paper we learned
about  a recent preprint \bakas\
where a class  of $D=4$  backgrounds with 2 Killing vectors
which solve the  one-loop equations of conformal invariance was considered.
The backgrounds (48),(49)  are of the type discussed  in \bakas.
The exact static  backgrounds considered in the  rest of  our paper
do not belong to the class of backgrounds ($
ds^2= f(t,z)(-dt^2+dz^2) +g_{ab}(t,z) dx^a dx^b , \ x^a=(x,y) $) studied  in
\bakas. }


\newsec{A  class of conformally invariant  sigma models}

\subsec{\bf  Solutions of conformal invariance conditions}
 The models we have  discussed above
 represent  exact conformal sigma  models belonging to
the  the  general class  of \sms (17),(26),   i.e.  ($ i=1,...,n$)
$$  S = {1\ov \pi \a'} \int d^2 z
 \big[   \del x^i \bd x_i
   +   F(x)   \del u \bd v  \big]   +  {1\ov 4 \pi  }
  \int d^2 z \sqrt {g^{(2)}}  R^{(2)} \p (x)   \  .    \eq{50} $$
In addition to the boost symmetry  in $u,v$ plane (which guarantees the
stability of the \sm (50) under renormalisation)
 these models are invariant under
the infinite-dimensional global symmetry $u'=u + f(\bar z), \ v'=v+ h(z)$
(these transformations are real on Minkowski world sheet)
corresponding to two conserved chiral currents.
It is straightforward to  find the conditions of  Weyl invariance  of the \sm
(50) by computing the corresponding  Weyl anomaly coefficients or $\bar
\beta$-functions.
The  leading-order  equations \call
 $$R_{mn} - \fourth H_{mpq} H_n^{\ pq} + 2D_m D_n \p=0\ , \ \  \  -\ha D_n
H^n_{\ pq} + \del_n  \p  H^n_{\ pq} =0 \ , $$
$$ {1\ov 6 \a' } (D-26) - \ha  D^2 \p + (\del \p)^2 - {1\ov 24} H_{mnk} H^{mnk}
=0 \ ,  \ \  \ D= n + 2 \ ,  $$
imply
the following conditions on the two functions $F$ and $\p$ in (17)
$$  -  \del_i \del_j h  +  \del_i \del_j \p = 0  \ , \ \ - \del^i\del_i h  + 2
\del_i \p \del^i h  =0\ ,  \ \ \ \ \ F \equiv   {\rm e}^{2 h(x) } \ ,  \eq{51}
  $$
$$  {26-D \ov 6 \a' }= -  \ha \del^i \del_i \p  +  \del^i \p \del_i \p  -\del^i
h \del_i \p  +   \del^i h\del_i h =   \del^i (h -\p) \del_i (h-\p)  \ . \eq{52}
 $$
As a consequence,
$$   \p -h\equiv \tilde \p  = \p_0 +  \r_i x^i   \ , \ \  \   \del^i \del_i  h
= 2\del_i h \del^i h
+ 2 \r_i \del^i h  \ , \ \     {D-26 }=  -   6  \a' \r_i \r^i\ ,  \eq{53} $$
where  $\r_i$ is an arbitrary constant vector.
The equation for $h$  is equivalent to\
$$  \del^i \del_i  \tilde F    =   2 \r_i \del^i \tilde F   \ ,  \ \
  \ \  \ \  \tilde F \equiv  F\inv = {\rm e}^{-2h}
\  .   \eq{54}
 $$
A linear combination of
solutions  of (54)   is\foot{
Let us note also that (54) can be put
 into the form of a massive equation  in flat  $n$-dimensional Euclidean  space
$  \  - \del^i\del_i  \TF'  + 2  |\r|^2   \TF'=0  ,  \ \  \TF \equiv {\rm
e}^{ \r_ix^i }  \TF' .  \
 $
When $\r=0$ we get $\TF = b_0 + b_ix^i$. this solution corresponds to a
non-trivial curved space which is dual (see Sect.4.3) to the  flat space.}
$$ \TF = b_0 + b_ix^i +  \sum_{r=1}^N \ep_r
{\rm e}^{ \a_r \cdot x } \ , \ \ \
 b_i \r^i=0 \ , \ \ \
\a_r \cdot x \equiv  \a_{ri}x^i ,   \eq{55}   $$
where $\ep_r$ are  arbitrary  constants and  a set of vectors  $\a_r$  should
satisfy (for each value of the index $r$ and  for a  fixed vector $\r_i$)
$$  |\a_r|^2 = 2 \r \cdot \a_r \ . \ \eq{56} $$
A particular   solution  of (54),(56)
is  found by taking  the number $N$ of the exponential terms in (55)
to be equal to  the number $n$  of coordinates $x^i$
 and identifying  $\a_r$ with the simple roots of  a  semisimple rank $n$  Lie
algebra and
 $\r_i$ with of the components of the vector equal to one half of the  sum of
all positive roots,
 or, equivalently, to the sum of the fundamental weights,
$$ \r = \ha \sum_{s=1}^m \a_s = \sum_{i=1}^n m_i \ , \ \  \ \
m_i\cdot \a_j =  \ha |\a_i|^2 \d_{ij} \ , \ \ \  \r\cdot \a_i =  \ha |\a_i|^2 \
. \eq{57} $$
 As a result,
$$ F= \big(  b_0 + b_ix^i +  \sum_{i=1}^n \ep_i {\rm e}^{ \a_i \cdot x })\inv
\ , \ \ \  \   \p =\p_0 +  \r_i x^i  - \ha \ln F   \ . \   \eq{58}  $$
We  have  thus checked that the model (26)  and, in particular,  all
$D=4$ backgrounds  (28),(29) obtained by gauging  the nilpotent   subalgebras
of maximally non-compact  $n=2$ algebras $sl(3,R), \ so(2,3)=sp(4), \ so(2,2)$
and $G_2$
 correspond to the particular cases of the
 solution (58) (with $b_0=b_i=0$).  The same conclusion is true for the
backgrounds  (47),(48),(49)
obtained by null gauging  of $SO(1,3)$.

The exact expression for the central charge of these models
is found from (53) taking $ \a'= 2/\k, \ \k= k-\ha c_G $
$$  C= n + 2 + 6 \a' |\r|^2 = n + 2 + { \ha c_G  d \ov k- \ha c_G} =
{  k d \ov k- \ha c_G}  -  d + n + 2   \ , \ \ \  \ 24 |\r|^2 = c_G d \ ,
\eq{59} $$
i.e. is given by the central charge of the original WZNW model minus the number
of
 degrees of freedom that have been gauged away.

The background fields in (26)  derived  by the
null gauging of the  WZNW models represent exact solutions of the conformal
invariance conditions
 (to all orders in $\a'$ in a particular scheme).
It is  natural to conjecture that,  in general, there exists a scheme in which
the backgrounds of the type
(50)
 satisfying  the one-loop conformal invariance  condition   (54)  are
actually  the solutions
 to all orders in $\a'$.  In fact, it is possible,
 to give a  path integral argument which demonstrates
this \hots.   Similar statement
  is known   for the  plane wave - type
backgrounds with one covariantly constant  null Killing vector    \amkl\hst\
and also for  the WZNW models or group spaces.
This is  certainly true for the model (67) related to (50) by  a duality
transformation as discussed below.
We shall also  note  that the model (50) has  two chiral currents, suggesting a
possibility of a Sugawara-type construction.

\subsec{\bf Relation to Toda model}

The equation  (54) can be interpreted as a condition of conformal invariance
for a massless (dimension two) perturbation in a linear dilaton background
and is very similar to the
corresponding one for the tachyon (dimension zero) perturbation $T(x)$
  $$  -  \del^i \del_i  T    +   2  \del_i \p_1  \del^i  T  -  2\k T =0 \ ,  \
\  \  \p_1  = \p_0 +       \s_i x^i  \ ,  \ \  \ \   2/\a'= \k=k-\ha c_G\ ,
$$
$$ \del^i \del_i  T    =  2  \s_i \del^i  T  -  2\k  T \ .     \eq{60}  $$
The  potential of the Toda model
$$ T(x) =  \sum_{i=1}^n \ep_i {\rm e}^{  \a_i \cdot x } \ , \    \eq{61}   $$
is a particular solution  of (60)  if
$$  |\a_i|^2 = 2 \s\cdot \a_i - 2\k  \ . \eq{62} $$
Eq. (62)  is  satisfied if (cf.(57))\foot{Similar  derivation of the Toda model
from the `$\b$-function'  conditions was given (in the simply-laced case) in
\tsey\ and in  a general non-abelian Toda model case
in  \jackk. The central charge of the Toda theory is thus  (cf. (53))
$C= n +  6 \a' |\s|^2 = n + 12 \k\inv  |\r + 2\k \d |^2$,  in agreement with
\oraif\jack. }
$$ \s  = \sum_{i=1}^n  m_i   +    2\k \sum_{i=1}^n {2\ov  |\a_i|^2} m_i \equiv
\r   +  2\k \d \ . \eq{63} $$
Comparing   (59) with  (54) or (62) with (56)  we see that the origin  of the
second $\d$ -
term in the Toda model dilaton \wip\oraif\jack\  is in the presence of  the
`classical
dimension' or `mass'
  term  $2\k T$ in (59) which absent in the case of the
sigma model - type interaction.

To  demonstrate the  equivalence  to the Toda model at the classical  level
 let us consider the
classical equations for the model  (50) (on a flat $2d$ background)
$$  \del\bd x_i  -\ha \del_i F \del u \bd v =0 \ , \ \ \
\del( F \bd v) =0 \ , \ \ \ \bd (F \del u) =0 \ .  \eq{64} $$
The model thus has two chiral currents. Integrating the last two equations and
substituting the
solutions in the first one  we get
$$ \del\bd x_i  + \ha  \chi \del_i F\inv =0 \ , \ \ \  F \bd v = \n (\bar z)  \
, \  \ F \del u = \m (z)
\ , \  \  \chi \equiv
\n (\bar z) \m (z) \  . \eq{65} $$
Since $\n, \m$ are chiral and $\chi$  has  a factorised form  they can   be
made constant by  the
conformal transformations of $z$ and $\bar z$ (the  \sm  (50) is always
conformally invariant at the classical level).
Equivalently, this can be  considered as a  gauge choice (alternative to the
light cone gauge), i.e. $u= a(\tau + \s), \ v= b(\tau-\sigma)$,   for the
conformal symmetry.
The  equation  for $x^i$ can be derived from the action
$$  S = {1\ov \pi \a'} \int d^2 z
 \big[   \del x^i \bd x_i
    -  \chi  F\inv  (x)  \big]  \  .    \eq{ 66 } $$
With $F$ given by   (21) (i.e. $F\inv = T= \sum_{i=1}^n \ep_i {\rm e}^{  \a_i
\cdot x }$)
 and  in the conformal gauge with   constant $\chi$
the  equation  for $x^i$ (65) and the action  (66) are  exactly
those of the Toda model.
This observation is, of course,  related to the derivation of the Toda model
from constrained WZNW model in \oraif.

As a consequence, the equations of the classical string propagation (including
the constraints)
on the backgrounds  (26) discussed  in the present  paper   are    exactly
integrable  since  their solutions can be directly expressed  in terms of
 the Toda model
solutions.

\subsec{\bf Dual sigma models }
Another   interesting  property of the model (50)  is that  the dual model
obtained by the standard  abelian
 duality transformation  in the $w= \ha (u+v)$  direction
has a  covariantly constant null Killing vector, i.e.
has   a  `plane wave' - type  structure
  (with the metric and dilaton  depending only  on  transverse  coordinates but
not on a  light-cone coordinate).
This is  a consequence of  the fact that the $uv$-component of the metric in
(50) is equal to the corresponding component of the antisymmetric tensor (what
is also the reason for the existence of the two chiral currents).
 Following the standard steps \busc\giveon\
of gauging the symmetry $u'=u + a, \ v'= v + a$,  i.e. adding the gauge field
strength term with a  Lagrange multiplier $ \tu$
and integrating out the gauge field,
we find for the dual model ($t= \ha (u-v)\equiv \tv$)
 $$  \tilde S = {1\ov \pi \a'} \int d^2 z
 \big[   \del x^i \bd x_i   + \TF(x) \del \tu \bd \tu   -  2 \del \tv \bd \tu
   \big]   +  {1\ov 4 \pi  }  \int d^2 z \sqrt {g^{(2)}}  R^{(2)} \tp (x) \   ,
   \eq{67} $$
$$ \TF= F\inv (x) \ , \ \ \ \ \ \tp = \p (x)  -  \ha \ln F (x)   \ . \eq{68} $$
The dual  space-time  metric   has one  covariantly constant null
Killing  vector while  the  vanishing antisymmetric tensor is zero.
Since the `transverse' part of the metric is flat,
the conditions of  the  Weyl invariance of the  model  (67)  are  given  (to
all orders in $\a'$)
  by  the `one-loop' conditions\foot{See,   e.g.,
\tsnul\ for a  general discussion.  Since the dilaton is assumed to depend on
the  transverse coordinates,
this model  is a  generalisation of the original `plane-wave' models
considered in \guv\amkl\hst\rudd.  Some special  cases, in particular,  the
$D=3$  case   of such model
 -- the duality rotation  of the $SL(2,R)$ WZNW model -- were   already
discussed
  (in connection with extremal black strings) in \horhorst\horwel (see  also
\ind;  note that a `nilpotent duality' discussed in  \ind\  in the $SL(2,R)$
case  is  just the standard abelian duality
if expressed in terms of the appropriate  coordinates (30),(31)).}
$$ -   \del^i \del_i   \TF  +  2\del_i \tp \del^i \TF
   = 0  \ , \ \ \  \del_i \del_j   \tp = 0 \ ,  \eq{69} $$
$$   {26-D \ov 6 \a' }=  - \ha  \del_i \del^i
  \tp   +  \del_i \tp \del^i \tp  \
   .     \   $$
This system of equations is equivalent to the one in the original model
(51),(52) (see (53),(54))
with  a solution already given in (55),(53),(58)
$$  \TF=   b_0 + b_ix^i +  \sum_{i=1}^n \ep_i {\rm e}^{ \a_i \cdot x }
\ , \ \ \  \   \tp =\p_0 +  \r_i x^i    \ , \   \eq{70}  $$
where $b_i$, $\a_i$ and $\r$ should satisfy the conditions in (55),(56).
The backgrounds (58) and (70) thus provide another example  of exact solutions
related by the standard (leading-order) duality (cf. \klts).
The classical equations of motion of the two dual models are, of course,  also
equivalent and  can be represented in the Toda-like  form
(65).

There remains an open question if
one can consider the isometric coordinate (with respect to which the duality is
performed) to be periodic so that the two conformal field theories can be
identified
following the idea of \rocver. This  could open a possibility for a proof of
unitarity of the models considered in this paper since the unitarity
of the `plane wave' model (67) is obvious in  the light cone gauge.

\newsec{Concluding remarks}

In this paper we have found a new class of conformal  \sms
which have Toda model - like structure  and  can be obtained by
gauging of  `null' (or nilpotent)  subgroups in   WZNW theories based on
non-compact groups.  The corresponding target space fields represent exact
string solutions.
Their possible physical interpretation remains an open question.
An interesting property  of these solutions is that the classical string
propagation is essentially described by the Toda model equations.

Let us  briefly comment on the geometrical properties of the  backgrounds
discussed above.
Computing the scalar curvature of the metric in (50) we find\foot{
It should be noted that even though  the string `feels' a combination of the
metric $and$ the antisymmetric tensor (so that the geometrical properties of
these backgrounds  are better reflected
in the string equations of motion) a point-like tachyonic state still follows
the geodesics
of the metric.}
$$  R= - 2 F\inv \del^i\del_i F  + \ha F^{-2}  \del^i F \del_i F =
  2 F \del^i\del_i F\inv - {7\ov 2}   F^{2}  \del^i F\inv \del_i F\inv  \ .
\eq{71}     $$
For $F$ in (26) or in (58) (with $b_0=b_i=0$) we obtain
$$ R = { \sum_{i,j=1}^n  ( 4 |\a_i|^2 - 7 \a_i\cdot \a_j )
\ep_i\ep_j {\rm e}^{ (\a_i + \a_j)\cdot x }
\ov  2 \sum_{i,j=1}^n \ep_i\ep_j
{\rm e}^{ (\a_i + \a_j)\cdot x } }\  .
 \eq{72}  $$
For comparison, in the case of the dual background (67)  all
 scalar curvature invariants vanish
while the  $\tu\tu$-component of the Ricci tensor is given by
$$ \tilde R_{\tu\tu} =
  - \ha  \del^i\del_i \TF =    - \ha  \del^i\del_i F\inv =  -\ha   \sum_{i=1}^n
\ep_i|\a_i|^2 {\rm e}^{ \a_i \cdot x } \ .  \eq{73} $$
Eq.(72)  gives  $R=-6$ in the simplest case of $\ep_2=...=\ep_n=0$ when the
model
is equivalent to the WZNW model for $SL(2,R) \times R^{n-1}$.
It is clear that the curvature  (72)  is non-singular if $F\inv$ cannot vanish,
i.e. if all $\ep_i$ have the same sign. For example, in the case of the four
dimensional   $ SO(2,2)$ background  (34), i.e.
  $F\inv =  {   {\rm e}^{\sqrt 2 x} + \ep   {\rm e}^{\sqrt 2\m y }  } , $
we have explicitly
$$  R =  { -3 {\rm e}^{2\sqrt 2 x}  + 4  \ep (1 + \m)    {\rm e}^{ \sqrt 2 x +
\sqrt 2\m y }
- 3  \m^2 \ep^2   {\rm e}^{2\sqrt 2\m y }
\ov ({\rm e}^{\sqrt 2 x} + \ep   {\rm e}^{\sqrt 2\m y } )^2 }   \ . \eq{74}  $$
The curvature is regular for $\ep = +1$,  approaching  a constant
  value   at large $x$ and $y$.
Similar expressions are found for other metrics in Section 3.

It  may be possible to construct new  solutions by
factorising over some discrete subgroups (i.e. by
 making some identifications of coordinates)  using as a motivation a
relation
\horwel\kala\ between the $3D$ black hole  \ban\  and $SL(2,R)$ WZNW model.


\newsec{Acknowledgements }
\noindent

C.K. is grateful to the Isaac Newton Institute, Cambridge
for hospitality while part of this work was done
and  acknowledges the support of the grants  GAUK 291 and 318.
A.A.T.  would like to thank G. Horowitz,  A.M. Semikhatov and  K. Sfetsos for
useful
comments. The work of A.A.T.   is  supported by   PPARC.


\vfill\eject
\listrefs
\end